\documentstyle[12pt,epic,eepic,epsf,rotate]{article} 
{}
\begin{document}
{\Large Computer Simulations of Pedestrian Dynamics\\[3mm]
and Trail Formation}\\[1cm]
Dirk Helbing (Stuttgart), P\'{e}ter Moln\'{a}r (Stuttgart), 
Frank Schweitzer (Berlin)\\
\vfill
Summary:\\
A simulation model for the dynamic behaviour of pedestrian crowds 
is mathematically formulated in
terms of a social force model, that means, pedestrians 
behave in a way as if they would be subject to
an acceleration force and to repulsive forces describing 
the reaction to borders and other pedestrians.
The computational simulations presented yield many 
realistic results that can be compared with video
films of pedestrian crowds. Especially, they show 
the self-organization of collective behavioural
patterns.
\par
By assuming that pedestrians tend to choose routes that 
are frequently taken the above model can be
extended to an active walker model of trail 
formation. The topological structure of the evolving trail
network will depend on the disadvantage of building 
new trails and the durability of existing trails.
Computer simulations of trail formation 
indicate to be a valuable tool for designing
systems of ways which satisfy the needs of pedestrians best.
An example is given for a non-directed trail 
network.\\[6mm] 
\clearpage

\subsubsection*{Introduction}

During the last two decades the investigation of 
dynamic pedestrian behaviour has found notable
interest among scientists from various disciplines 
for several reasons: First, architects and urban
planners need powerful tools for designing 
pedestrian areas, subway or railroad stations, entrance
halls, shopping malls, escape routes, etc. 
Second, there are some theoretical challenges like the
description of interaction effects (that can 
lead to jams, blockages) and of influences of the geometry
of pedestrian facilities. The observation of 
striking analogies with gases and fluids is of particular
interest for physicists. Third, pedestrian models 
can be experimentally tested, since all model
quantities like places and velocities of 
pedestrians are easily measurable. Empirical data material of
flow measurements and film material already 
exist. Fourth, pedestrian models can give valuable hints
for the development of more general or other 
behavioural models. For example, there exist analogies
with models of opinion formation [1,2].
\par
The historical modelling approaches to pedestrian dynamics 
were quite different. Traffic engineers
usually try to fit simple {\em regression models} to 
flow measurements [3]. However, since these
models do not describe interaction effects in an 
adequate way, these flow measurements and
regression analyses have to be performed for every 
special situation. In particular, it is not possible to
exactly predict the pedestrian flows in pedestrian areas 
or buildings with an extraordinary
architecture during the planning phase.
There have also been suggested some {\em queuing models} [4,5].
However, these are only suitable for certain kinds of 
questions, since they do not explicitly take into
account the effects of the concrete geometry of pedestrian 
facilities. Furthermore, {\sc Borgers} and
{\sc Timmermans} [6] have developed a model for the 
{\em route choice behaviour} of pedestrians
in dependence of their demands, the city entry points, and 
the store locations. However, this
approach does not model the effects of pedestrian 
interactions, in spite of the fact that pedestrians
will take detours if their prefered way is crowded.
In the 1970s, {\sc Henderson} suggested that pedestrian crowds 
can be described by the {\sc Navier-Stokes} equations of 
{\em fluid-dynamics} [7]. However, his approach
implicitly assumes energy and momentum to be collisional 
invariants which is obviously not the case
for pedestrian interactions. Therefore, in the 1990s a 
fluid-dynamic pedestrian model has been
derived on the basis of a pedestrian 
specific {\em gaskinetic} ({\sc Boltzmann}-like) model [8,1].
This model starts from observations of individual 
pedestrian behaviour and takes into account the
intentions, desired velocities and pair interactions 
of the considered pedestrians. 
\par
Apart from analytical investigations, for practical applications 
a direct simulation of individual
pedestrian behaviour is favourable, since a 
numerical solution of the fluid-dynamic equations is very
difficult. As a consequence, current research 
focuses on the microsimulation of pedestrian crowds. In
this connection, a {\em social force model} of 
pedestrian dynamics has been suggested [9,10] which
is related to {\em molecular dynamics} [11]. 
A simple forerunner of this kind of model has been
proposed by {\sc Gipps} and {\sc Marksj\"o} [12].
\par
The social force model can be extended to an {\em active walker 
model} [13--16] by inclusion of a {\em trail formation mechanism}, which has
originally been suggested for the description of trunk trail formation by ants
[16]. One interesting application of this active walker model
is the construction of way systems that are optimal compromises between 
{\em minimal way systems} (that necessitate many detours) 
and {\em direct way systems} (that are very expensive and 
need too much space) [17].

\subsubsection*{The social force model of pedestrian dynamics}

Pedestrians are {\em used} to the situations 
they are normally confronted with. Their behaviour is
determined by their experience which 
reaction to a certain stimulus (situation) will be the {\em best}.
Therefore, their reactions are usually rather 
{\em `automatic'} and well predictable. The
corresponding behavioural rules can be put into an 
{\em equation of motion}. According to this
equation, changes of the {\em actual velocity} $\vec{v}_\alpha(t)$ 
of a pedestrian $\alpha$ are given
by a vectorial quantity $\vec{f}_\alpha(t)$: 
\begin{equation}
\frac{d\vec{v}_\alpha}{dt} = \vec{f}_\alpha(t) + \mbox{\em fluctuations.}
\label{Eq1}
\end{equation}
$\vec{f}_\alpha(t)$ can be interpreted as {\em social force} 
that describes the influence of
environment and other pedestrians on the 
individual behaviour. However, the social force is not {\em
exerted} on a pedestrian. It rather describes the concrete 
{\em motivation to act}. The {\em
fluctuation term} takes into account random variations of 
the behaviour, which can arise, for example, in
situations where two or more behavioural 
alternatives are equivalent, or by accidental or deliberate
deviations from the usual rules of motion.
\par
There are some major differences between social forces and 
forces in physics. First, for social forces
the {\sc Newton}ian law {\em actio = reactio} does not hold. 
Second, energy and momentum are
{\em no collisional invariants} which implies that there is 
no energy or momentum conservation.
Third, pedestrians (or, more general, individuals) are {\em 
active systems}, that produce forces and
perform changes themselves. Fourth, instead of by {\em 
momentum transfer} via virtual particles the
effect of social forces comes about by {\em 
information exchange} via complex mental, psychological
and physical processes.
\par
In the following we will specify the social force model of pedestrian motion:
\begin{itemize}
\item[1.] A pedestrian $\alpha$ wants to walk into a 
{\em desired direction} $\vec{e}_\alpha$ (the
direction of his/her next destination) with a 
certain {\em desired speed} $v_\alpha^0$.
The desired speeds of pedestrians are almost normally 
distributed: \begin{equation}
P(v^0) = \frac{1}{\sqrt{2\pi\sigma}}\mbox{e}^{-(v^0 - \langle v^0 \rangle)^2
 / (2\sigma)} \, .
\end{equation}
A deviation of the {\em actual velocity} $\vec{v}_\alpha$ from 
the {\em desired velocity}
$\vec{v}_\alpha^{\,0} := v_\alpha^0 \vec{e}_\alpha$ 
leads to a tendency $\vec{f}_\alpha^{\,0}$
to approach $\vec{v}_\alpha^{\,0}$ again within 
a certain {\em relaxation time} $\tau_\alpha$. This
can be described by an {\em acceleration term} of the form
\begin{equation}
 \vec{f}_\alpha^{\,0}(\vec{v}_\alpha,v_\alpha^0\vec{e}_\alpha)
:= \frac{1}{\tau_\alpha} (v_\alpha^0 \vec{e}_\alpha - \vec{v}_\alpha ) \, .
\label{Eq3}
\end{equation}
$\vec{f}_\alpha^{\,0}$ is a measure for the {\em 
`pressure of time'}, that means the {\em motivation
to get ahead}. 
\item[2.] Pedestrians keep a certain distance
to {\em borders} (of buildings, walls, streets,
obstacles, etc.). This effect $\vec{f}_{\alpha B}$
can be described by a repulsive, monotonic decreasing potential $V_{B}$:
\begin{equation}
 \vec{f}_{\alpha B} (\vec{r}_{\alpha} - \vec{r}_B^{\,\alpha})
 = - \nabla_{\vec{r}_{\alpha}}
 V_{B} (\|\vec{r}_{\alpha} - \vec{r}_B^{\,\alpha}\|) \, .
\label{Eq4}
\end{equation}
Here, $\vec{r}_\alpha$ is the actual {\em location} 
of pedestrian $\alpha$. In addition,
$\vec{r}_B^{\,\alpha}$ denotes the location of the 
piece of the border that is nearest to location
$\vec{r}_\alpha$. \item[3.] The motion of a 
pedestrian $\alpha$ is influenced by other pedestrians
$\beta$. These interactions cause him/her to 
perform {\em avoidance manoeuvres} or to slow down
in order to keep a situation dependent 
distance to other pedestrians. A quite realistic description of
pedestrian interactions results from the assumption 
that each pedestrian respects
the {\em `privat spheres'} of other 
pedestrians $\beta$. These {\em territorial effects}
$\vec{f}_{\alpha\beta}$ can be modelled by 
{\em repulsive potentials} $V_{\beta}(b)$:
\begin{equation}
 \vec{f}_{\alpha\beta}(\vec{r}_{\alpha} - 
\vec{r}_{\beta}) = - \nabla_{\vec{r}_{\alpha}} V_{\beta}
[b(\vec{r}_{\alpha} - \vec{r}_{\beta})] \, . \label {Eq5}
\end{equation}
\par
The sum over the repulsive potentials $V_{\beta}$ defines
the {\em interaction potential} which
influences the behaviour of each pedestrian:
\begin{equation}
 V_{\rm int}(\vec{r},t) := \sum_\beta V_{\beta} \{ b [ \vec{r}
 - \vec{r}_\beta(t) ] \} \, .
\label{Eq6}
\end{equation}
For $V_{\beta}(b)$ we will assume a monotonic decreasing function 
in $b$ with equipotential lines
having the form of an ellipse that is 
directed into the direction of motion (see Fig. 1). The reason for
this is that a pedestrian requires space 
for the next step, which is taken into account by other
pedestrians. $b$ denotes the {\em semi-minor 
axis} of the ellipse and is given by
\begin{equation}  2b = \sqrt{ (\|\vec{r}_{\alpha} - \vec{r}_{\beta}\|
 + \|\vec{r}_{\alpha} - \vec{r}_{\beta} - v_\beta
\, \Delta t \, \vec{e}_\beta \| )^2 - ( v_\beta \, \Delta t )^2  } \, .
\label{Eq7}
\end{equation}
$s_\beta
:= v_\beta \, \Delta t$ is about the 
{\em step width} of pedestrian $\beta$. \par
\parbox[b]{7.5cm}{
\unitlength0.5cm
\begin{center}
\begin{picture}(6.5,5.5)(0,0.5)
\thicklines
\put(0.5,1){\vector(1,0){6}}
\put(6,0.3){\makebox(0,0){{$b$}}}
\put(1,0.5){\vector(0,1){6}}
\put(-0.5,6){\makebox(0,0){{$V_{\beta}(b)$}}}
\spline(1,6)(2.1,2.95)(3.15,1.8)
\spline(3.15,1.8)(3.95,1.25)(6,1.03)
\end{picture}
\end{center}}
\hfill
\parbox[b]{8cm}{                                                        
\unitlength0.75cm
\begin{center}
\begin{picture}(5,3.5)(2,1.5)                                  
\thicklines
\put(3,3){\circle*{0.3}}
\put(3,3){\vector(1,0){3}} \put(3.1,2.5){\makebox(0,0){{$\vec{r}_\beta$}}}
\put(5.2,2.5){\makebox(0,0){{$v_\beta \, \Delta t \, \vec{e}_\beta$}}}
\put(4.5,3){\vector(0,1){2.5}}
\put(5,5.25){\makebox(0,0){{$b$}}}
\put(4.5,3){\ellipse{3.16}{1}}
\put(4.5,3){\ellipse{3.61}{2}}
\put(4.5,3){\ellipse{4.24}{3}}
\put(4.5,3){\ellipse{5}{4}}                                 
\end{picture}
\end{center}}
{\small{\bf Fig. 1:} Left: Decrease of the 
repulsive interaction potential $V_\beta(b)$ with $b$.\\
Right: Elliptical equipotential
lines of $V_{\beta} [b(\vec{r} - \vec{r}_{\beta})]$ for 
different values of $b$.}
\end{itemize}
Since all the effects 1 to 3 influence a pedestrian's decision 
at the {\em same} moment, we shall
assume that their total effect is given by 
the {\em sum} of all effects like this is the case for forces.
The {\em social force (total motivation)} $\vec{f}_\alpha$ 
is, therefore, given by
\begin{eqnarray}
 \vec{f}_\alpha(t)
 &:=&  \vec{f}_\alpha^{\,0}(\vec{v}_\alpha,
 v_\alpha^0\vec{e}_\alpha)
 + \vec{f}_{\alpha B}(\vec{r}_\alpha - \vec{r}_B^{\,\alpha})
 + \sum_{\beta(\ne \alpha)} \vec{f}_{\alpha \beta} (
 \vec{r}_\alpha - \vec{r}_\beta) \nonumber \\
 &=&  \frac{1}{\tau_\alpha} (v_\alpha^0 \vec{e}_\alpha - \vec{v}_\alpha )
 - \nabla_{\vec{r}_{\alpha}} \Big[ V_B(\|\vec{r}_\alpha 
 - \vec{r}_B^{\,\alpha} \|) + V_{\rm int}(\vec{r}_\alpha,t) \Big] \, .
\label{Eq8}
\end{eqnarray}
According to (\ref{Eq1}), $\vec{f}_\alpha(t)$ determines 
the temporal change of the actual velocity
$\vec{v}_\alpha(t)$. Together with the equation
\begin{equation}
 \frac{d\vec{r}_\alpha}{dt} := v_\alpha(t) \, ,
\label{Eq9}
\end{equation}
(\ref{Eq1}) and (\ref{Eq8}) characterize the {\em social force model} 
of pedestrian motion. The corresponding model of
pedestrian {\em crowds} has the form of {\em nonlinearly 
coupled stochastic differential equations}.
\par
The social force model (\ref{Eq1}), (\ref{Eq8}), (\ref{Eq9}) of 
pedestrian dynamics has been simulated on a
computer for a large number of interacting pedestrians 
confronted with different situations. Despite
the fact that the proposed model is very simple 
it describes a lot of observed phenomena very
realistically. For example, under certain 
conditions the {\em emergence} of new spatio-temporal 
patterns of collective behaviour can be observed:
\begin{itemize}
\item[1.] the development of lanes (groups) 
consisting of pedestrians walking into the {\em same}
direction (see Fig. 2),
\item[2.] oscillatory changes of the walking direction 
at narrow passages (for example, doors) (see Fig. 3),
\item[3.] the spontaneous formation of roundabout 
traffic at intersections (see Fig. 4).
\end{itemize}
This formation of new spatio-temporal patterns is 
called {\em `self-organization'} since it is not
caused by the special {\em initial} or 
{\em boundary conditions}, but by the {\em non-linear
interactions} of pedestrians. Furthermore, the        
self-organization phenomena mentioned above are not
the effect of strategical considerations, because 
the pedestrians were assumed to behave in a rather
`automatic' way.
\begin{figure}[htbp]
\epsfysize=16.5cm 
\centerline{\rotate[r]{\hbox{\epsffile[193 33 406 810]{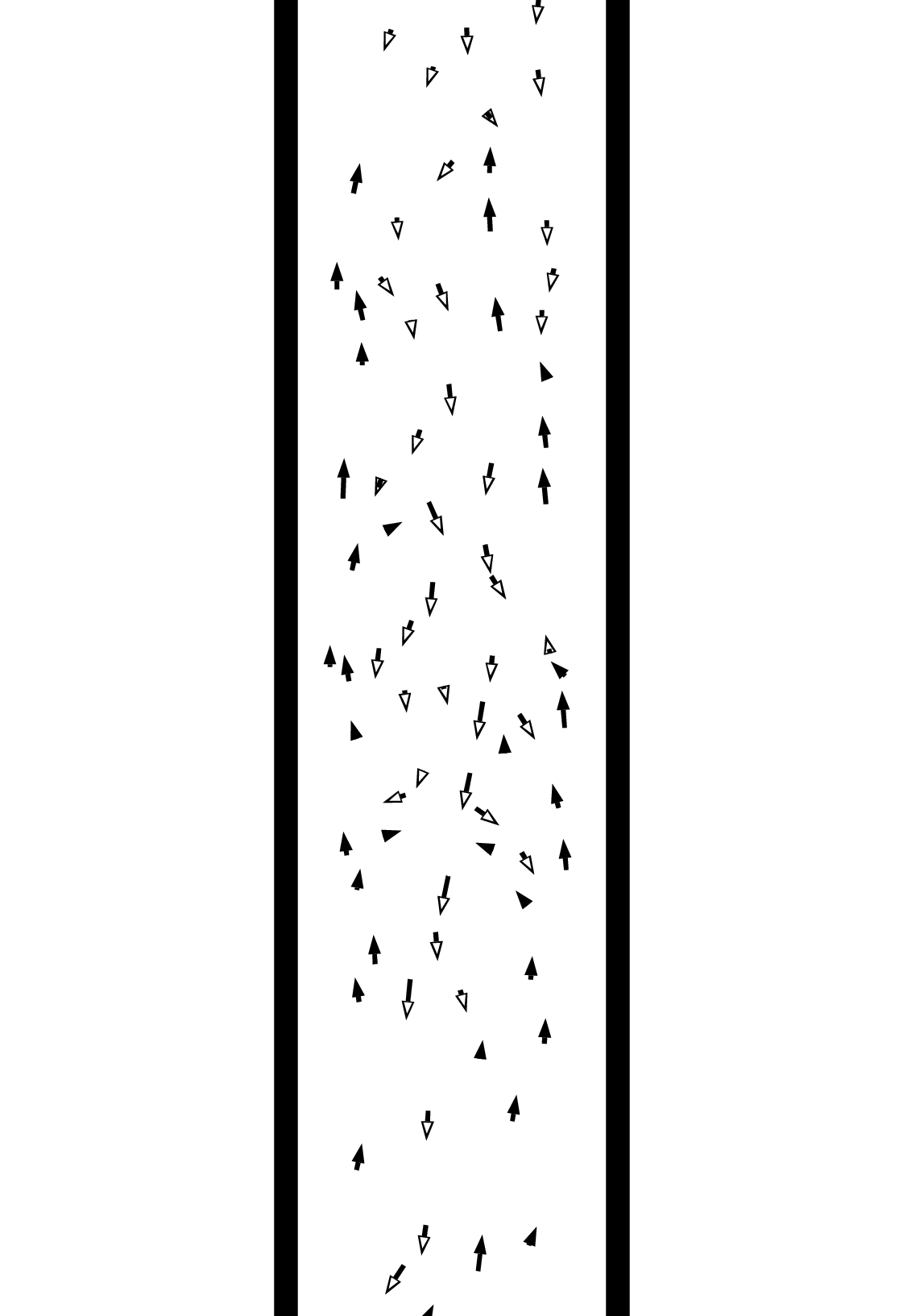}}}}
{\mbox{ }\\[-2mm]
\small{\bf Fig. 2:} Above a critical pedestrian density one finds the
formation of lanes consisting of pedestrians with an uniform walking
direction. Here, the computational result shows three lanes. 
Black arrows represent pedestrians who desire
to walk from left to right, white arrows pedestrians who want to move into the
opposite direction. The lengths of the arrows indicate the speeds of the
corresponding pedestrians.}
\end{figure}

\subsubsection*{Trail formation of pedestrians}

In the following, we will focus our interest on the self-organization 
of {\em systems of ways}, which
has been described recently by means of an {\em active walker model} [16]. 
The term {\em `active walker'} [13--16]
is derived from {\em `random walker'} which means a particle that 
moves randomly in space due to
the effect of fluctuations. An active walker is {\em also} 
subject to fluctuations, but is additionally
able to perform certain actions:
\begin{itemize}
\item[1.] An active walker can change his/her/its 
environment by setting {\em markings}.
\item[2.] In addition, an active walker is able to 
{\em read} these markings which influence
his/her/its behaviour in a specific way.
\end{itemize}
\epsfxsize=6cm                                               
\centerline{\rotate[r]{\hbox{\epsffile[123 33 482 810]{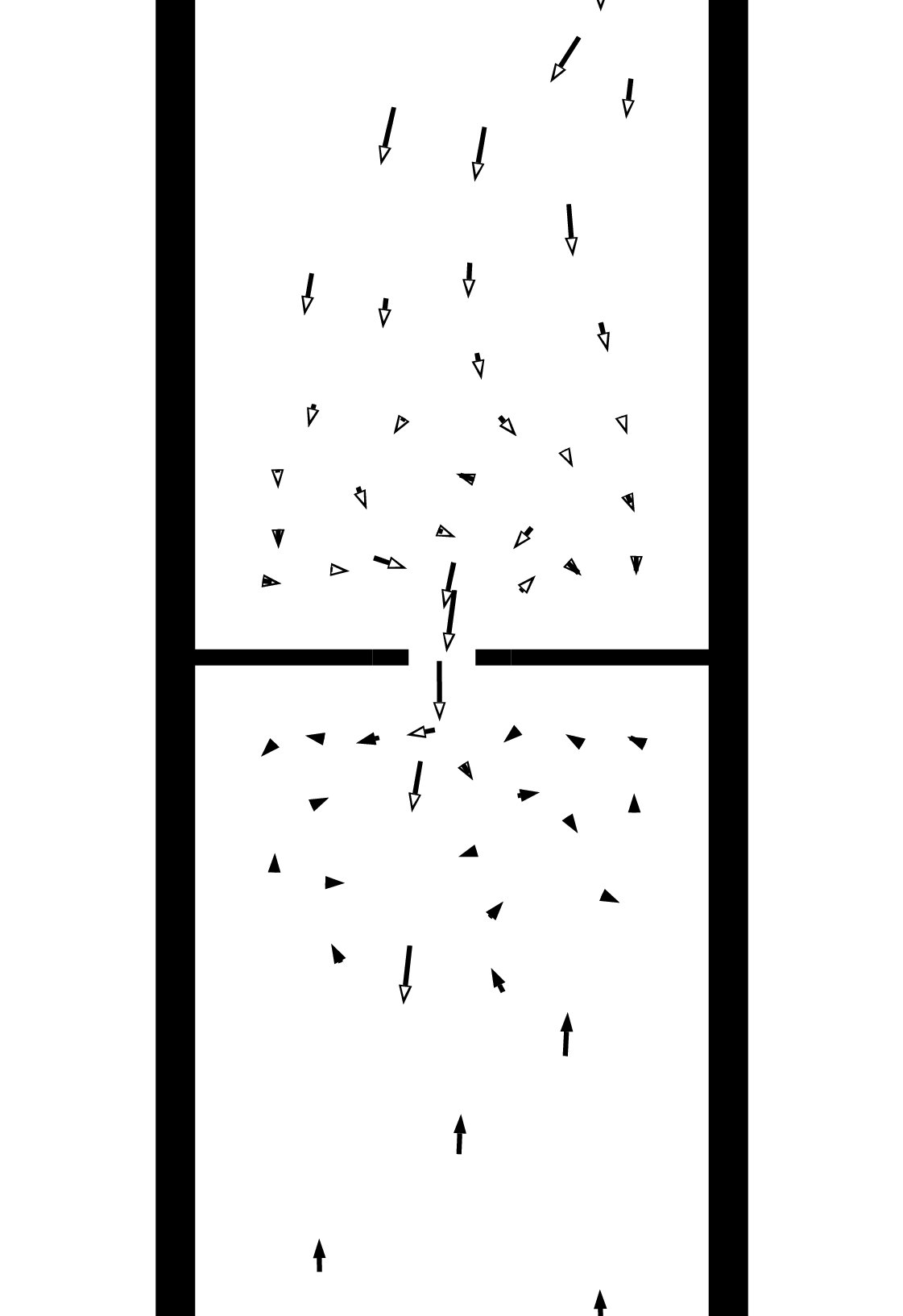}}}}
\mbox{ }\\
\epsfxsize=6cm 
\centerline{\rotate[r]{\hbox{\epsffile[123 33 482 810]{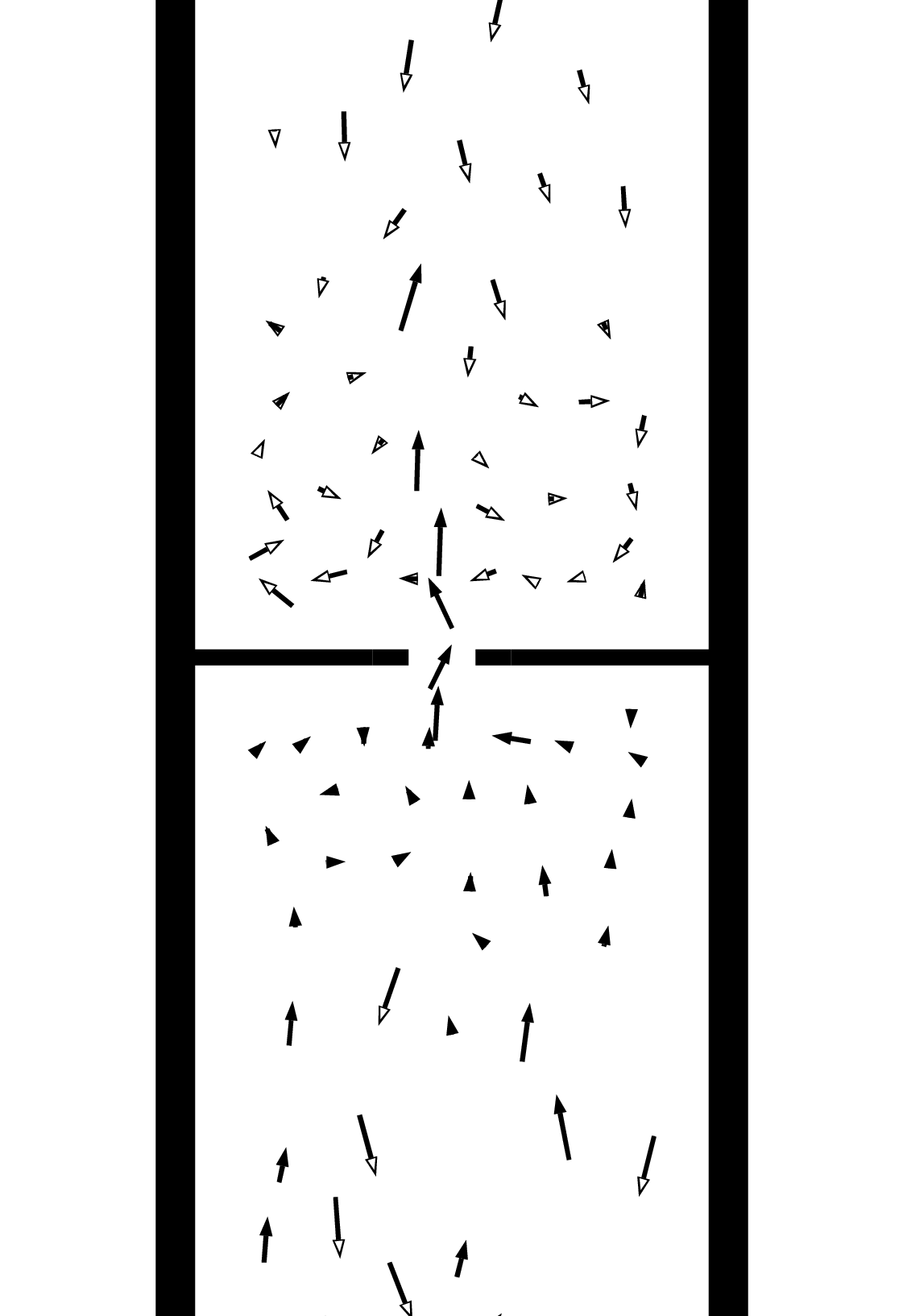}}}}
{\mbox{ }\\[-2mm]
\small{\bf Fig. 3:} Different moments of two pedestrian groups 
that try to pass a narrow door into opposite directions.
If one pedestrian has been 
able to pass a narrow door, other pedestrians
with the same desired walking direction can follow easily whereas
pedestrians with an opposite desired direction of motion have to wait
(above). After some time the pedestrian stream is stopped
by the pressure of the opposing group, and the door is subsequently
captured by pedestrians who intend to pass the door into the opposite
direction (below). The change of the passing direction
may occur several times.}
\par                                            
That means, active walkers {\em interact indirectly} with                       
each other via environmental changes
(markings). The markings have their own properties which 
depend on their special type. For
example, they may decay with time or diffuse spatially.
\par
The pedestrian dynamics discussed above can be expanded with 
a trail formation mechanism by                                   
assuming that each pedestrian produces a {\em trail}
consisting of {\em footprints} (which play the role of 
markings) [15,16]. These footprints will (exponentially)
decay in the course of time with a rate $1/T$ where $T$ means
the {\em durability} of the markings. The decay rate $1/T$ should not be too 
large since then the trails will almost have vanished 
before another pedestrian intends to take a similar route.
Introducing a {\em trail potential}
$V_{\rm tr}(\vec{r},t)$ which describes the additional effect 
$\vec{f}_{\rm tr}(\vec{r},t) := - \nabla_{\vec{r}} V_{\rm tr}(\vec{r},t)$
of the markings on pedestrian motion, the temporal
change of $V_{\rm tr}(\vec{r},t)$ is given by a {\em decay term} and 
terms $Q_\alpha(\vec{r},t)$ which reflect the {\em production} of new
footprints by the pedestrians:
\begin{equation}
 \frac{dV_{\rm tr}(\vec{r},t)}{dt} 
 = - \frac{1}{T} V_{\rm tr}(\vec{r},t) + \sum_\beta Q_\beta(\vec{r},t) \, .
\label{diffgl}
\end{equation}
\epsfxsize=10cm 
\centerline{\rotate[r]{\hbox{\epsffile[25 67 560 770]{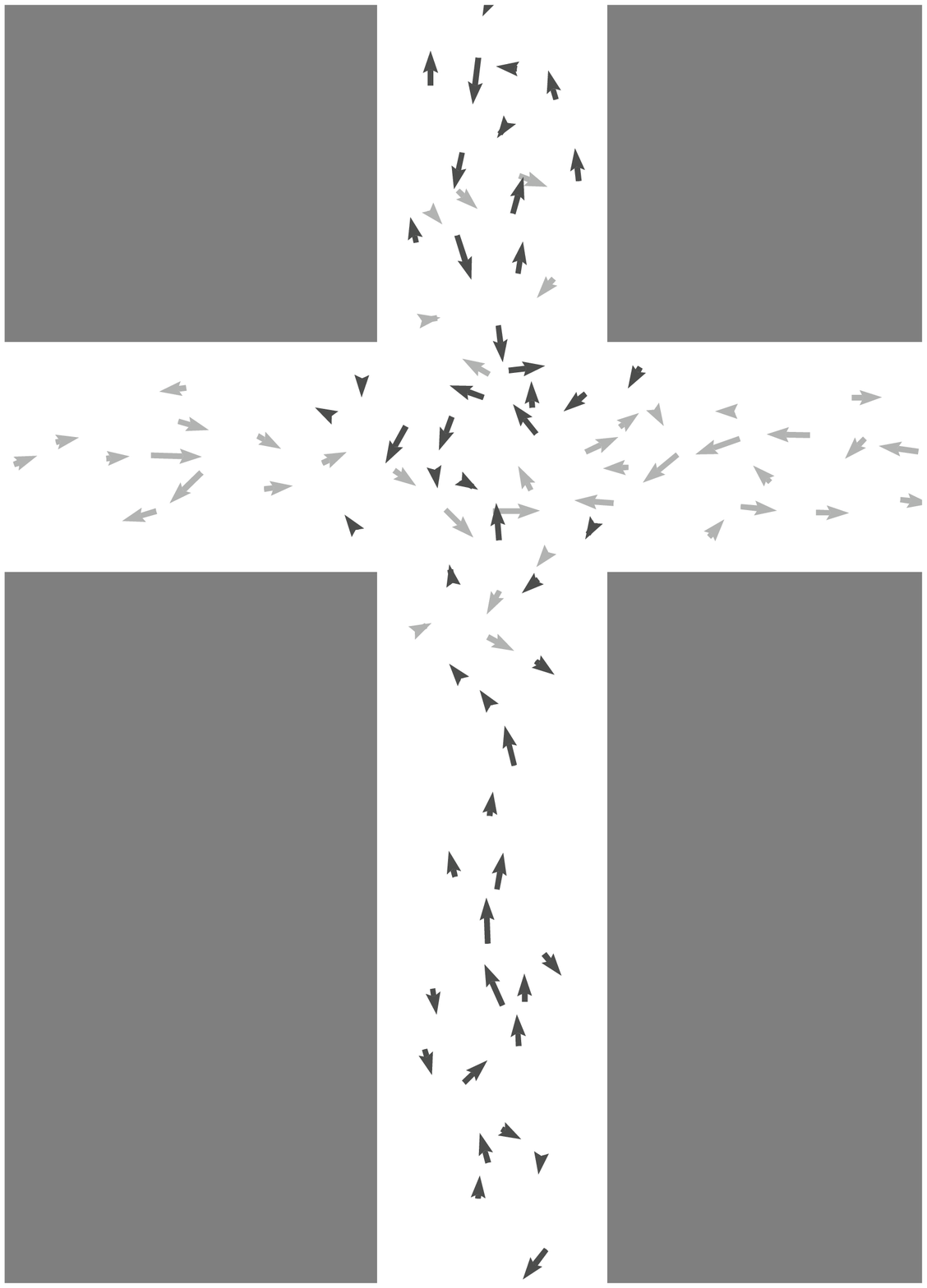}}}}
{\mbox{ }\\[-2mm]
\small{\bf Fig. 4:} At crowded intersections there develops
roundabout traffic for certain time periods. The related rotation direction
changes from time to time. White arrows represent pedestrians who desire
to walk in vertical direction, black arrows pedestrians who want to move in
horizontal direction.}
\par
Integration of (\ref{diffgl}) leads to
\begin{equation}
 V_{\rm tr}(\vec{r},t) = \int\limits_{t_0}^t dt' \, Q(\vec{r},t') 
 \exp \left( -\frac{t - t'}{T} \right)  \qquad \mbox{with} \qquad
 Q(\vec{r},t) := \sum_\beta Q_\beta(\vec{r},t) \, .
\label{intdar}
\end{equation}
The {\em production terms} $Q_\alpha(\vec{r},t)$ 
are modelled by {\em attractive 
potentials}, since it should be easier (more convenient) for
pedestrians to take already existing trails than to clear new 
ways. For reasons of simplicity we will assume that $Q_\alpha(\vec{r},t)$
has the same functional form as the repulsive potential
$V_{\alpha}\{b[\vec{r} - \vec{r}_{\alpha}(t)]\}$, but with an opposite sign.
If the parameter $q$ reflects the strength of
the attraction of new markings, we then have
\begin{equation}
 Q_\alpha(\vec{r},t) = - q V_\alpha \{ b [ \vec{r} - \vec{r}_\alpha(t)]\}
 \qquad \mbox{and} \qquad 
 Q(\vec{r},t) = - q V_{\rm int}(\vec{r},t) \, . 
\label{prod}
\end{equation}
$q$ can be interpreted as the {\em advantage of using 
existing trails}. If $q$ is too small, no trail
formation will be observed. On the other hand, if 
$q$ is too large, the attraction effect of trails will
destroy the repulsive effect of pedestrian interactions 
which would lead to collisions of pedestrians.
A trail formation, of course, is observed only if the trails are reinforced
by regular usage.
\par
In our simulations we use a
triangular grid for the two-dimensional space and 
a discretization of time. The related discretization
of (\ref{intdar}) reads then:
\begin{equation}
 V_{\rm tr}(\vec{r},t) = \Delta t \! \sum_{n=1}^\infty Q(\vec{r},t - n \,
\Delta t) \exp \left( - \frac{n \, \Delta t}{T} \right) \, .
\end{equation}
Moreover, the equation of motion for a pedestrian $\alpha$
obtains the form
\begin{equation}
  \frac{\vec{v}_\alpha(t+\Delta t) - \vec{v}_\alpha(t)}{\Delta t} =
 \frac{1}{\tau_\alpha} [v_\alpha^0 \vec{e}_\alpha - \vec{v}_\alpha(t) ]
 - \nabla_{\vec{r}_{\alpha}} V_{\rm tot}(\vec{r}_\alpha,t) 
 + \mbox{\em fluctuations}  
\end{equation}
where we have introduced the {\em total potential} 
\begin{equation}
  V_{\rm tot}(\vec{r},t) = V_{B}(\|\vec{r} - \vec{r}_B \|)
 + V_{\rm int}(\vec{r},t) + V_{\rm tr}(\vec{r},t)  \, .
\end{equation}  
Computer simulations of the above described trail formation model
are expected to be a useful tool for designing efficient systems of ways
that satisfy the pedestrians' requirements. For
this purpose one has to specify the advantage $q$ of using 
existing trails and to simulate the
expected flows of pedestrians that enter the 
considered system at certain entry points with the
intention to reach certain destinations. 
According to the above described model, after some time a
way system will evolve which takes into account 
the pedestrians' route choice habits.
For the simulation of real situations the corresponding
results may serve as suitable planning guidelines for 
architects or urban planners.

\subsubsection*{Formation of non-directed trail systems}

The social force model of pedestrian motion described above assumes 
of each pedestrian certain intentions like a desired direction, 
desired speed, and desired distance from obstacles or other pedestrians. 
Whereas these assumptions allow us to describe the interactions of
`real' pedestrians, we now want to focus on the {\em basic} features of the
process of trail formation itself. Therefore, we drop some of the assumptions
above and simplify the model. The pedestrians are 
assumed to be {\em simple} walkers which 
only make {\em local} decisions about
the direction of their next step. In particular, these walkers have 
{\em no memory} about the way they have gone or about certain destinations.
\par
During their walk, the walkers leave markings (`footprints') again.
If we neglect the repulsive interactions between the walkers and assume 
no obstacles in the system, the only force on the walkers is given by 
the trail potential $V_{\rm tr}(\vec{r},t)$. The equation of motion 
of walker $\alpha$ is now given by equation (\ref{Eq9}) and
\begin{equation}
 \frac{d\vec{v}_\alpha}{dt} = - \gamma \vec{v}_\alpha 
 - \nabla_{\vec{r}_\alpha} V_{\rm tr}(\vec{r}_\alpha,t) + 
 \sqrt{2 \gamma^2 D}\xi_\alpha(t) 
\label{frank1}
\end{equation}
where we have added a dissipative {\em `friction term'} $-\gamma\vec{v}_\alpha$
which describes a loss of energy. $\gamma$ is the {\em friction
coefficient} of the walkers. The last term of (\ref{frank1}) 
specifies the fluctuation term of equation (\ref{Eq1}) 
as a {\em white random force (white noise)}, where the strength 
of the fluctuations is related to the {\em diffusion coefficient} $D$.
It is well known that equations (\ref{Eq9}) and (\ref{frank1}) can, in
the {\sc Smoluchowski} limit [18], be reduced to
\begin{equation}
 \frac{d\vec{r}_\alpha}{dt} = - \frac{1}{\gamma} \nabla_{\vec{r}_\alpha}
 V_{\rm tr}(\vec{r}_\alpha,t) + \sqrt{2D} \xi_\alpha(t) 
\label{frank2}
\end{equation}
which corresponds to $d\vec{v}_\alpha/dt \approx 0$.
The first part of (\ref{frank2}) describes the fact that the walkers 
tend to follow the {\em gradient} of the markings (that means, to move
into the direction of the {\em strongest} marking in their neighbourhood), 
whereas the second part represents the 
fluctuations that keep them moving otherwise with a certain probability. 
As long as the gradient of 
$V_{\rm tr}(\vec{r},t)$ is small and the fluctuations are 
large, the walkers behave
nearly as random walkers, but the action of the walkers (the production of
markings) can lead to a {\em supercritical value} of the attractive trail
potential which causes a {\em spatially restricted} motion [15].
\par
Equation (\ref{prod}) for the production of markings implies
that the footprints have an attraction potential  
with a certain spatial extension, 
expressed by the parameter $b$. This means a trail becomes already 
attractive in a certain distance from the trail itself. If we assume instead
that the attraction potential is sharply peaked around the markings, the
production term can be simplified by means of $\delta$-functions 
$\delta[\vec{r} - \vec{r}_\alpha(t)]$ at the
actual positions $\vec{r}_\alpha$ of the walkers:
\begin{equation}
 Q(\vec{r},t) := - q \sum_\beta \delta[(\vec{r} - \vec{r}_\beta(t) ] \, .
\end{equation}  
Furthermore, we suppose that the walkers are able to see a 
marked trail only in a certain range of space (a certain angle 
of view in the direction of their movement). That means they 
percept their environment similar to real pedestrians rather than to 
physical particles which `feel' a gradient in every direction. If the walkers
find markings within the perceived 
range of space, they can choose the direction of
their next step towards the strongest marking, but, due to the fluctuations,
 with a certain probability
they `ignore' the markings and make a random decision on their next step.
Provided that they follow the markings, the footprints are reinforced.
\par
As a result, a trail system will evolve (see Fig. 5) which consists of 
a number of major and minor trails. The frequency of use is 
coded with a grey scale. This trail system is non-directed, it does 
not connect any points of destination. Since the simulated surface is only 
{\em partially} covered with markings, the walkers really use the trails during
their movement instead of moving around anywhere.
\par
\begin{figure}[htbp]
\epsfxsize=9cm 
\centerline{\rotate[r]{\hbox{\epsffile[176 310 420 546]
 {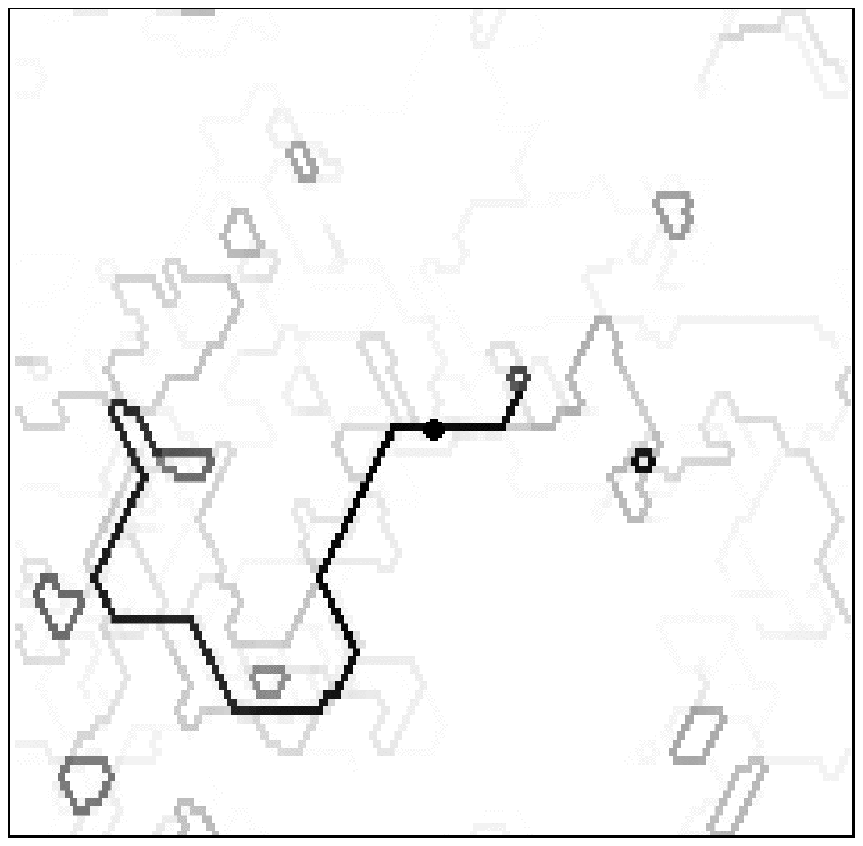}}}}
{\mbox{ }\\[-3mm]
\small{\bf Fig. 5:} 
Non-directed trail system originated by 100 active walkers after 
5000 time steps. (The starting point was in the middle 
of the triangular lattice of size 100x100.)}
\end{figure}
\par
The observed trail system is a {\em non-planned structure}. It
emerges from the {\em local interaction} of the walkers with 
the surface, which in turn changes the movement of the walkers. There 
is an interplay between the reinforcement of trails and the {\em selection}
of trails that are less used. The resulting structure finally consists
of only those trails which could be maintained by the walkers.
All other trails have vanished after a certain time.
\par
As indicated by (\ref{frank2}) the density of the trail system depends on 
the relation between the attractive and the dissipative effects. If 
the diffusion constant $D$ is small, only a few trails will evolve. 
The density of trails increases with $D$, and above a critical value
$D_{\rm crit}$ for $D$, a sharply distinguished trail system is no
longer observed (every walker creates its own trail, then) [19].

\subsubsection*{References:}

\begin{itemize}
\item[{[1]}] D. Helbing. {\em Stochastische Methoden, nichtlineare 
Dynamik und quantitative
Modelle sozialer Prozesse}. Shaker, Aachen, 1993.
\item[{[2]}] D. Helbing. A mathematical model for 
behavioral changes by pair interactions and its
relation to game theory. {\em Angewandte Sozialforschung} {\bf 18} 
(3), 117--132, 1993/94.
\item[{[3]}] {\em Highway Capacity Manual}, Chap. 13. 
Transportation Research Board, Special
Report 209, Washington, D.C., 1985.
\item[{[4]}] A. J. Mayne. Some further results in the 
theory of pedestrians and road traffic. {\em
Biometrica} {\bf 41}, 375--389, 1954.
\item[{[5]}] G. G. L{\o}v{\aa}s. Modelling and simulation of 
pedestrian traffic flow. In: {\it
Modelling and Simulation 1993. Proceedings of the 
1993 European Simulation Multiconference},
Lyon, France, June 7--9, 1993.
\item[{[6]}] A. Borgers and H. J. P. Timmermans. City centre 
entry points, store location patterns
and pedestrian route choice behaviour: A microlevel simulation 
model. {\it Socio-Economic Planning
Science} {\bf 20}, 25--31, 1986.
\item[{[7]}] L. F. Henderson. On the fluid mechanics of 
human crowd motion. {\it Transportation
Re\-search} {\bf 8}, 509--515, 1974.
\item[{[8]}] D. Helbing. A fluid-dynamic model for the 
movement of pedestrians. {\it Complex
Systems} {\bf 6}, 391--415, 1992.
\item[{[9]}] D. Helbing. A mathematical model for the 
behavior of pedestrians. {\it Behavioral
Science} {\bf 36}, 298--310, 1991.
\item[{[10]}] D. Helbing and P. Moln\'{a}r. A social force 
model for pedestrian dynamics.
Submitted to {\it Physical Review E}, 1994.
\item[{[11]}] W. G. Hoover. {\it Molecular Dynamics.} Lect. Notes Phys., Vol.
258. Springer, Berlin, 1986.
 \item[{[12]}] P. G. Gipps and B. Marksj\"o. A micro-simulation model 
for pedestrian flows. {\em
Mathematics and Computers in Simulation} {\bf 27}, 95--105, 1985.
\item[{[13]}] R. D. Freimuth and L. Lam. In: {\em Modeling Complex 
Phenomena}, edited by L.
Lam and V. Naroditsky. Springer, New York, 1992.
\item[{[14]}] D. R. Kayser, L. K. Aberle, R. D. 
Pochy and L. Lam. Active walker models: tracks
and landscapes. {\em Physica A} {\bf 191}, 17--24, 1992.
\item[{[15]}] F. Schweitzer and L. Schimansky-Geier. Clustering 
of ``active walkers'' in a two-component system. {\em Physica A} {\bf 206}, 
359--379, 1994.
\item[{[16]}] F. Schweitzer, K. Lao and F. Family. Active 
Walker simulate trunk trail formation
by ants. Submitted to {\em Adaptive Behavior}, 1994.
\item[{[17]}] E. Schaur. {\it Ungeplante Siedlungen/Non-Planned Settlements}.
Information of the Institute for Leightweight Structures (IL), University
of Stuttgart, No. 39.
\item[{[18]}] C.~W. Gardiner. {\it Handbook of Stochastic Methods}.
Springer, Berlin, 2nd edition, 1985.
\item[{[19]}] F. Schweitzer and L. Schimansky-Geier. Clustering of 
active walkers: Phase transition from local interactions. In: {\em Fluctuations 
and Order: The New Synthesis}, edited by M. Millonas. Springer, New York, 1994.
\end{itemize}
\end{document}